\documentclass[reprint,superscriptaddress,amsmath,amssymb,aps,prb,floatfix]{revtex4-2}
\usepackage{pdfpages,pgffor,graphicx,dcolumn,bm,xr,physics,chemformula,siunitx,color,soul,float,appendix, subfigure, booktabs,array, multirow}

\newcommand{\bmk}{{\bf k}}
\newcommand{\bmkk}{{\bf K}}
\newcommand{\bmq}{{\bf q}}
\newcommand{\bmqq}{{\bf Q}}
\newcommand{\bmr}{{\bf r}}

\newcommand{\td}{\ch{TiO2}}

\colorlet{blue}{blue!70!black} 
\colorlet{red}{red!60!black}

\makeatletter
\AtBeginDocument{\let\LS@rot\@undefined}
\makeatother

\begin{document}


\title{Identification of large polarons and exciton polarons \\ in rutile and anatase polymorphs of titanium dioxide}

\author{Zhenbang Dai}
\affiliation{Oden Institute for Computational Engineering and Sciences, The University of Texas at Austin, Austin, Texas 78712, USA}
\affiliation{Department of Physics, The University of Texas at Austin, Austin, Texas 78712, USA}
\author{Feliciano Giustino}%
\email{fgiustino@oden.utexas.edu}
\affiliation{Oden Institute for Computational Engineering and Sciences, The University of Texas at Austin, Austin, Texas 78712, USA}
\affiliation{Department of Physics, The University of Texas at Austin, Austin, Texas 78712, USA}
        
\date{\today}

\begin{abstract}
Titanium dioxide (\td) is a wide-gap semiconductor with numerous applications in photocatalysis, photovoltaics, and neuromorphic computing. 
The unique functional properties of this material critically depend on its ability to transport charge in the form of polarons, namely narrow electron wavepackets accompanied by local distortions of the crystal lattice. It is currently well established that the most important polymorphs of \td, the rutile and anatase phases, harbor small electron polarons and small hole polarons, respectively.
However, whether additional polaronic species exist in \td, and under which conditions, remain open questions.
Here, we provide definitive answers to these questions by exploring the rich landscape of polaron quasiparticles in \td\ via recently developed \textit{ab initio} techniques. In addition to the already known small polarons, we identify three novel species, namely a large hole polaron in rutile, a large quasi-two-dimensional electron polaron in anatase, and a large exciton polaron in anatase. These findings complete the puzzle on the polaorn physics of \td\, and pave the way for systematically probing and manipulating polarons in a broad class of complex oxides and quantum materials.
\end{abstract}
\maketitle

\begin{figure}
\includegraphics[width=1\columnwidth]{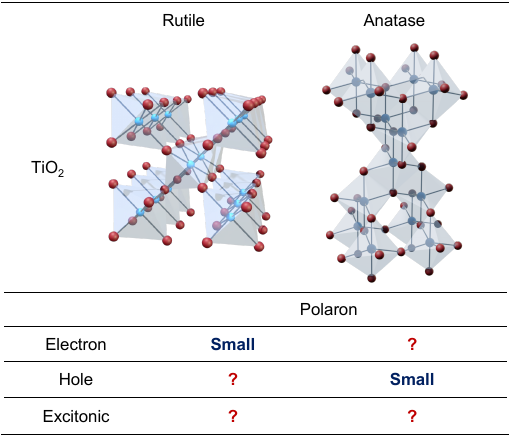}
\caption{
\label{fig:summary_of_polarons} \textbf{The polaron puzzle in \ch{TiO2}}. 
The consensus view is that rutile \ch{TiO2} hosts small electron polarons, while anatase hosts small hole polarons. The possible existence of hole polarons in rutile, electron polarons in anatase, and exciton polarons or self-trapped excitons in either polymorph remain open questions. 
}
\end{figure}

Titanium dioxide (\ch{TiO2}) is a uniquely versatile semiconductor with numerous technological applications. Since the discovery of the Fujishima-Honda water splitting reaction half a century ago~\cite{fujishima1972electrochemical}, \ch{TiO2} has remained the most widely used photocatalyst in artificial photosynthesis and environmental remediation~\cite{yang2008anatase, shen2023room, lee2019reversible,nakata2012tio2,morales2012design,gomes2011influence,selloni2008anatase,diebold2003surface,gong2006steps,iqbal2021charge,scanlon2013band}. 
In solar photovoltaics, \ch{TiO2} has been the most efficient photoanode for dye-sensitized solar cells since their inception~\cite{o1991low,atli2018multi,atilgan2022ni}, and is currently the most popular choice for the electron transport layer in perovskite solar cells~\cite{lee2012efficient,Kim2012}. In the context of microelectronics, the discovery of the memristive effect in \ch{TiO2} marked the beginning of neuromorphic computing~\cite{strukov2008missing}.

These applications crucially depend on the nature and transport properties of excess electrons and holes in \ch{TiO2}~\cite{luttrell2014anatase,schneider2014understanding}. 
Charge carriers in this material are known to exist in the form of polarons~\cite{elmaslmane2018first, kang2019first,deskins2007electron, deskins2009intrinsic, spreafico2014nature,labat2008structural, chiodo2010self, yang2010photoinduced, zhang2007niobium, macdonald2010situ, livraghi2014nature, tang1995urbach,gallart2018temperature}, which are composite quasiparticles consisting of localized electrons accompanied by distortions of the crystal lattice~\cite{franchini2021polarons}. Polarons are a direct manifestation of electron-phonon coupling~\cite{giustino2017electron,sio2019polarons,lafuente2022unified,lee2021facile,falletta2022many,dai2023explrn_prb,dai2024excitonic}.
At weak coupling, polarons behave like free electrons, except with heavier masses~\cite{alexandrov2010advances, feynman1955slow}; at strong coupling, polarons form localized wavepackets resembling atomic or molecular orbitals, and their transport occurs via thermally-activated hopping between lattice sites~\cite{emin2013polarons,holstein1959studies}. Beyond charge carriers, optical excitations in \ch{TiO2} are also strongly affected by electron-phonon coupling. For example, self-trapped excitons (STEs), which are bound electron-hole pairs coupled to localized lattice distortions, have been proposed to play an important role in the optical properties of this material~\cite{gallart2018temperature, sildos2000self, tang1995urbach,kernazhitsky2014room}.

Despite two decades of experimental and theoretical efforts, the nature of polarons and their excitonic counterparts in the most important polymorphs of titanium dioxide, rutile and anatase, remains a subject of intense debate. Currently, it is widely accepted that rutile harbors small electron polarons, whereas anatase hosts small hole polarons~\cite{elmaslmane2018first}. However, significant uncertainty persists regarding the potential existence of hole polarons in rutile, electron polarons in anatase, and STEs in either polymorph, see Fig.~\ref{fig:summary_of_polarons}.

\begin{figure*}
\includegraphics[width=0.98\textwidth]{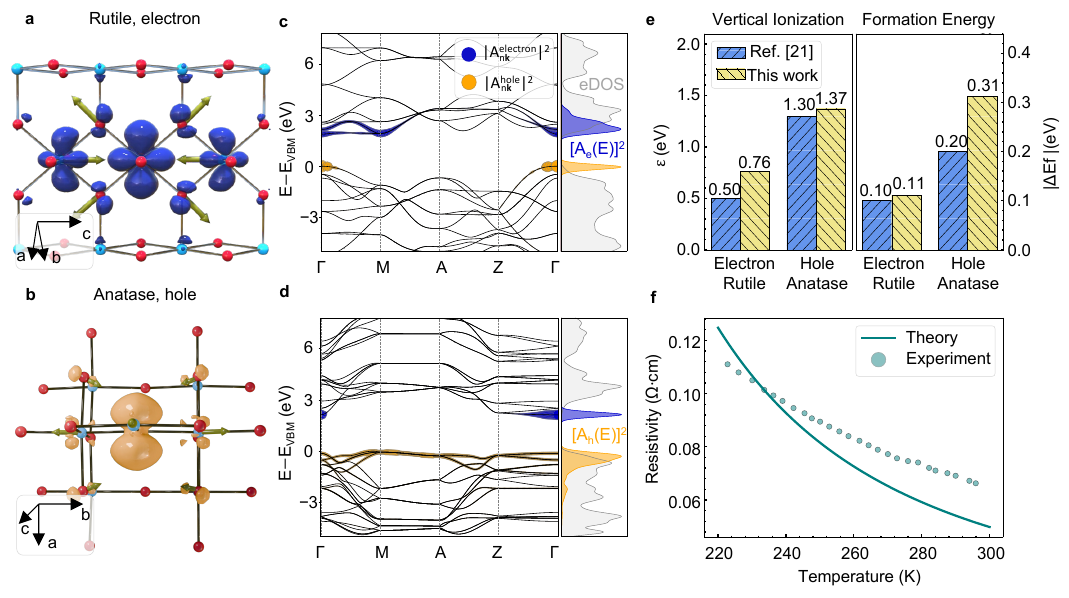}
\caption{
\label{fig:small_polarons} 
\textbf{Small polarons in rutile and anatase \ch{TiO2}}. 
\textbf{a} Charge density isosurface (blue isosurface) and atomic displacements (green arrows) of small electron polaron in rutile. Ti and O are in light blue and red, respectively. 
\textbf{b} Charge density isosurface (orange isosurface) and atomic displacements (green arrows) of small hole polaron in anatase.
\textbf{c} Contribution from Kohn-Sham states to the electron polaron (blue discs) and hole polaron (orange discs) in rutile, and corresponding electron spectral function $A^2(E)$ overlaid on the electron density of states (eDOS). 
\textbf{d} Contribution from Kohn-Sham states to the electron polaron (blue discs) and hole polaron (orange discs) in anatase and the corresponding electron spectral function $A^2(E)$ overlaid on the eDOS.
In the right panels of \textbf{c} and \textbf{d}, all spectral functions and eDOS are normalized so that the top of the range coincides with the highest peak in each case. 
\textbf{e} Comparison of polaron energetics between this work and previous results by DFT hybrid functional calculations~\cite{elmaslmane2018first, deak2012quantitative}.
{\color{black}The vertical ionization energy here is defined as the polaron eigenvalue [Eq.~(\ref{eqn:plrneqn})] of the polaronic structure referenced to the polaron eigenvalue of the perfect periodic structure.}.
\textbf{f} Resistivity of rutile \ch{TiO2} calculated using the Emin-Holstein-Austin-Mott theory (line), and comparison with experiments (disks)~\cite{zhang2007niobium}.
}
\end{figure*}

Small polarons in \ch{TiO2} have been investigated theoretically via density functional theory (DFT)~\cite{elmaslmane2018first, setvin2014direct, deskins2007electron,deskins2009intrinsic,de2022polaron,kokott2018first}, and their existence has been confirmed experimentally via electron paramagnetic resonance (EPR) spectroscopy~\cite{yang2013intrinsic, berger2005light} and transport measurements~\cite{zhang2007niobium}. In particular, in the case of \textit{n}-doped rutile,
the decrease of electrical resistivity with increasing temperature is the hallmark of small electron polaron hopping~\cite{zhang2007niobium}; whereas, for \textit{p}-doped rutile, EPR studies from different groups reached opposite conclusions on the possible presence of small hole polarons in this polymorph~\cite{yang2010photoinduced, macdonald2010situ}. 
In the case of \textit{n}-doped anatase, the measured resistivity vs.\ temperature curve follows the standard Bloch-Gr\"uneisen law~\cite{zhang2007niobium}, ruling out small electron polarons. 
However, these findings are at odds with some DFT investigations reporting small electron polarons in anatase~\cite{deskins2007electron,spreafico2014nature, kim2017dissimilar}. 
Establishing whether these polymorphs host large polarons is even more challenging: On the one hand, distinguishing large polarons from free electrons is difficult in experiments as these particles carry similar fingerprints; On the other hand, calculations of large polarons have remained beyond reach until very recently, owing to prohibitive computational costs~\cite{sio2019ab, elmaslmane2018first} and the uncertainty introduced by the DFT self-interaction error~\cite{setvin2014direct, deskins2007electron, deskins2009intrinsic, elmaslmane2018first}. A similar uncertainty persists for STEs in \ch{TiO2}: Some experiments reported exciton self-trapping in both the rutile and anatase phases, while others found that only anatase harbors such excitations~\cite{gallart2018temperature, sildos2000self, tang1995urbach,kernazhitsky2014room,baldini2017strongly}. 

Here, we provide the missing pieces of the puzzle by presenting evidence for large hole polarons in rutile \ch{TiO2}, large electron polarons in anatase \ch{TiO2}, and large exciton polarons in anatase. To this end, we deploy recently-developed state-of-the-art \textit{ab initio} methods for computing small and large polarons, exciton polarons, self-trapped excitons~\cite{sio2019ab, lafuente2022ab, dai2023explrn_prb}, as well as their transport properties~\cite{ponce2018towards, ponce2020first, leveillee2023ab}. This methodology was shown to overcome the aforementioned limitations of DFT-based approaches~\cite{sio2019ab,lafuente2022ab,dai2023explrn_prb}. The computational setup is described in the Methods section and in SI Appendix, Supplemental Note~1, and a synoptic view of polaron properties is provided in SI Appendix, Table~S1.

\begin{figure*}
\centering
\includegraphics[width=0.8\textwidth]{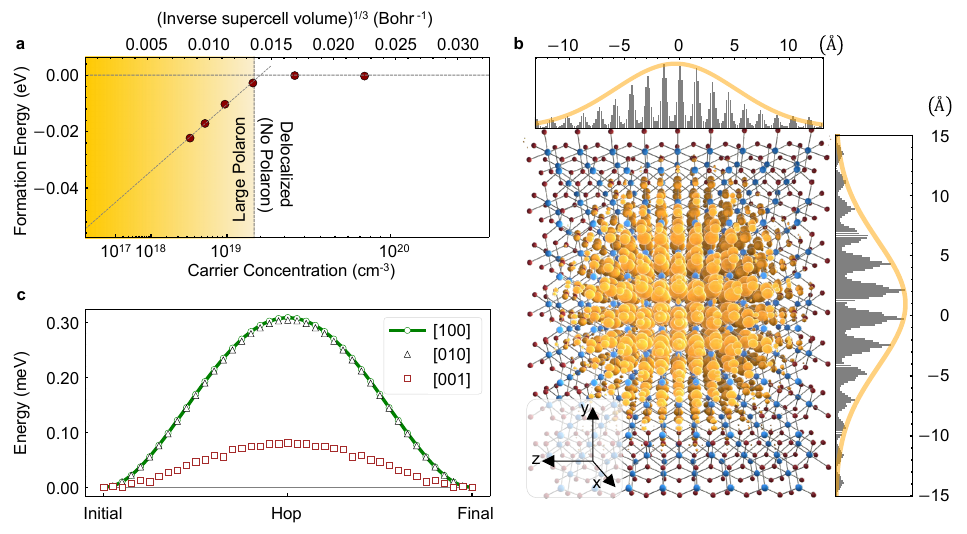}
\caption{
\label{fig:large_h_polarons} 
\textbf{Large hole polaron in rutile \ch{TiO2}}.
\textbf{a} Formation energy of the hole polaron in rutile \ch{TiO2} vs.\ carrier concentration. The latter is determined by the supercell volume, which is controlled by the Brillouin zone sampling in the calculations. 
Localization occurs when below the Mott density $\mathrm{1.7\!\times\!10^{19}~cm^{-3}}$.
The data points are plotted according to the formation energies and the inverse supercell lengths, and the carrier concentrations are computed separately for each supercell size.
\textbf{b} Charge density isosurface (orange isosurface) of hole polaron in rutile. 
Ti and O atoms are in light blue and red, respectively. The upper panel and side panel display the planar averages of the charge density (grey sticks); orange contours are obtained by fitting the charge densities to a Gaussian function. 
\textbf{c} Hopping barriers for the hole polaron in rutile \ch{TiO2}, for various directions. 
}
\end{figure*}

\section*{Small polarons}
Given the broad consensus on the existence of small electron polarons in rutile and small hole polarons in anatase~\cite{elmaslmane2018first}, we investigate these species to validate our methodology against experiments and hybrid functional calculations~\cite{elmaslmane2018first}.
Figs.~\ref{fig:small_polarons}(a) and \ref{fig:small_polarons}(b) show our calculated small polarons (electron in rutile, hole in anatase).
The orbital character and localization length of these polarons agree quantitatively with prior hybrid functional investigations~\cite{elmaslmane2018first, morita2023models}, underscoring the reliability of our approach. 
In the case of rutile, the electron polaron consists of three $t_{\rm 2g}$ Ti-$3d$ orbitals in adjacent unit cells along the $c$ axis [Fig.~\ref{fig:small_polarons}(a)]. The corresponding lattice distortion consists of Ti$^{4+}$ displacements toward the center of the polaron, and O$^{2-}$ displacements away from it, as expected from electrostatics. This three-site polaron is analogous to the trimeron that was identified in connection to the Verwey metal-insulator transition in magnetite \ch{Fe3O4}~\cite{baldini2020discovery,senn2012charge}. 
In the case of anatase, the hole polaron consists of a single O-$2p$ orbital [Fig.~\ref{fig:small_polarons}(b)], with Ti$^{4+}$ moving away from the polaron center and second nearest-neighbor O$^{2-}$ moving toward it. The formation and character of these small polarons can be rationalized by inspecting the band edges in Figs.~\ref{fig:small_polarons}(c) and (d).
In both cases, we see non-dispersive bands connecting multiple valleys, and the polarons consist of coherent superpositions of Kohn-Sham states throughout the Brillouin zone. This delocalization in reciprocal space is consistent with the strong localization of the polarons in real space. In addition, the polaron weights are seen to concentrate near the band edges, so that the character of frontier orbitals is clearly reflected in the polaron wavefunctions (compare Fig.~\ref{fig:small_polarons} with Supplemental Fig.~S1).

We obtain polaron formation energies, which correspond to the adiabatic stabilization energy of the polaron with respect to a delocalized excess electron or hole, of 111\,meV for electrons in rutile, and 312\,meV for holes in anatase. 
These values are in line with previous DFT calculations using hybrid functionals \cite{deak2012quantitative, elmaslmane2018first}, as shown in Fig.~\ref{fig:small_polarons}(e). We ascribe small differences between the two approaches to the fact that hybrid functional calculations suffer from a residual self-interaction error, which is completely eliminated in our method~\cite{sio2019ab,lafuente2022ab}.

\begin{figure*}
\includegraphics[width=1\textwidth]{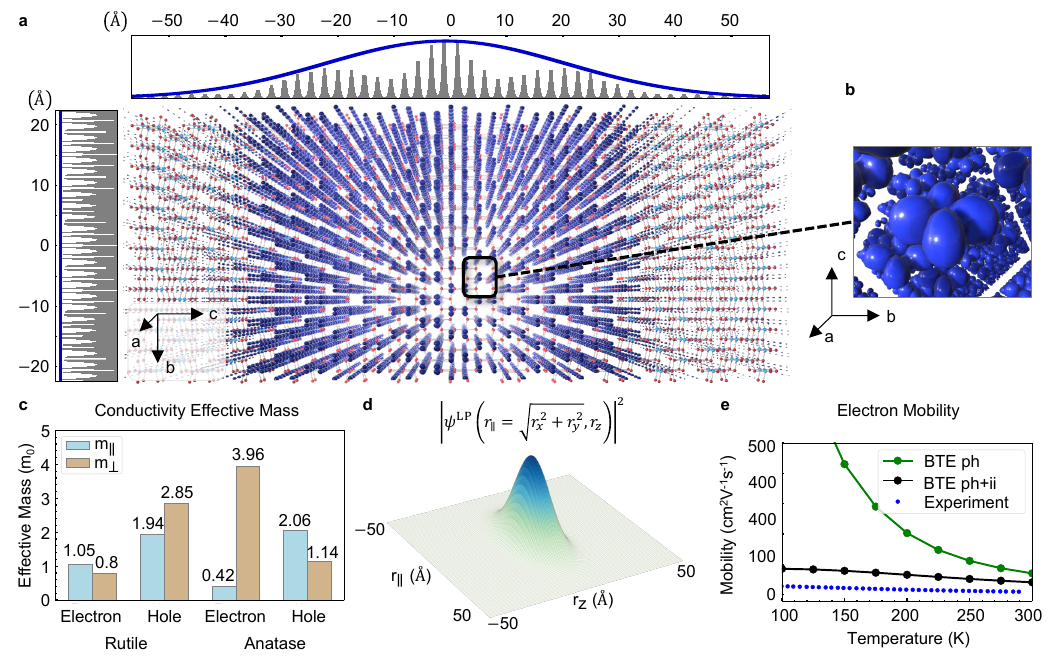}
\caption{
\label{fig:2d_e_polarons} 
\textbf{Large quasi-2D electron polaron in anatase \ch{TiO2}}. 
\textbf{a} Charge density isosurface (blue isosurface) of electron polaron in anatase. The upper and side panels show the planar averages of the polaron charge densities (grey sticks), and the blue contours are obtained by fitting the densities to Gaussian functions.
Ti and O atoms are in light blue and red, respectively. 
\textbf{b} Close-up view of the polaron charge density near a Ti atom, which takes the shape of a $d_{xy}$ orbital. Atoms and bonds are removed for clarity. 
\textbf{c} Summary of the conductivity effective masses for rutile and anatase, showing significant anisotropy between the $ab$ plane ($\parallel$) and the $c$ axis ($\perp$).
\textbf{d} Polaron wavefunction $\psi^{\rm LP}$ obtained from the anisotropic Landau-Pekar model, showing that the polaron size in the $ab$ plane is much larger than the size of the supercell used in \textbf{a}.
\textbf{e} Electron mobility of anatase \ch{TiO2}, as calculated via the \textit{ab initio} Boltzmann transport equation, including both phonon scattering (BTE ph) and ionized-impurity scattering (BTE ph+ii). Calculations are compared to the experimental data of Ref.~\citenum{zhang2007niobium}. Scattering by ionized Nb impurities is important to quantitatively describe experimental data, yielding a room-temperature mobility of  40~$\mathrm{cm^2 V^{-1} s^{-1}}$, in good agreement with the experimental value 10~$\mathrm{cm^2 V^{-1} s^{-1}}$~\cite{zhang2007niobium}. The impurity concentration, 1.25$\times$10$^{21}$~cm$^{-3}$, is taken from experiments.
}
\end{figure*}

To make connection with experiments on $n$-doped rutile, we investigate the energetics of polaron hopping between nearest-neighbor Ti sites (SI Appendix, Fig.~S2). We find a barrier of 13\,meV, which matches closely the barrier determined from EPR measurements in nominally undoped rutile \ch{TiO2}, 24$\pm$5\,meV~\cite{yang2013intrinsic}.
This barrier can be used to calculate the temperature-dependent resistivity via the Emin-Holstein-Austin-Mott theory~\cite{deskins2007electron, morita2023models}. We consider only nearest-neighbor hopping as the barrier for second nearest-neighbor hopping is an order of magnitude higher (SI Appendix, Fig.~S2), and use the carrier concentration from Ref.~\citenum{zhang2007niobium}. Fig.~\ref{fig:small_polarons}(f) shows that our calculated resistivity is in very good agreement with experiments, lending theoretical support to the notion that electron transport in rutile mainly occurs via small polaron hopping (see SI Appendix, Supplemental Note~2 for calculations of hopping mobility). We are not aware of similar experimental data for the small hole polarons in anatase.

\begin{figure*}
\includegraphics[width=0.98\textwidth]{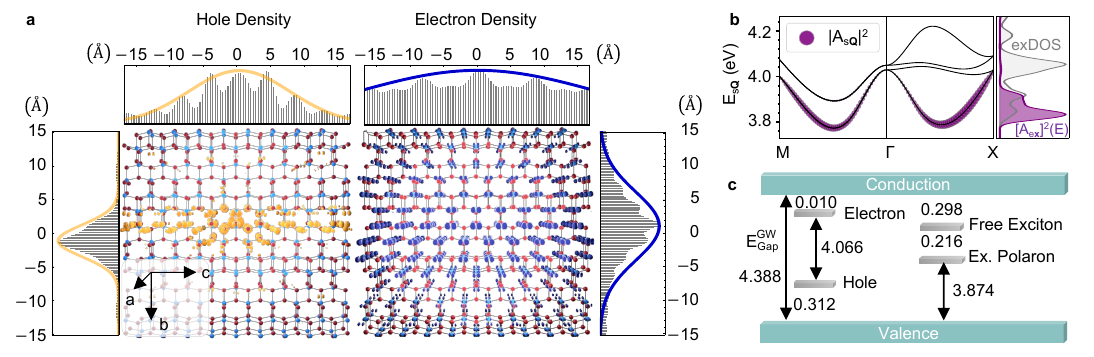}
\caption{
\label{fig:exciton_polarons} 
\textbf{Exciton polaron in anatase \ch{TiO2}}. \textbf{a} Three leftmost panels: Charge density isosurface (orange isosurface) for the hole density of the exciton polaron in anatase. The top and side panels show the planar averages of the charge density (sticks) and the orange contours are Gaussian fits.
Three rightmost panels: Charge density isosurface (blue isosurface) for the electron density of the exciton polaron in anatase. The top and side panels show the planar averages of the charge density (sticks) and the blue contours are Gaussian fits. Ti and O atoms are in light blue and red, respectively.
\textbf{b} Exciton band structure (black lines) in anatase and contributions of each BSE eigenstate to the exciton polaron (purple discs). The side panel shows the corresponding exciton spectral function $A^2(E)$ superimposed to the exciton density of states (exDOS). The lattice contributions to the exciton polaron are shown in SI Appendix, Fig.~S3(e). \textbf{c} Schematic energy diagram of charged and neutral excitations in anatase \ch{TiO2} (all values in eV). The formation energies of charged polarons (10\,meV and 312\,meV for electron and hole, respectively) correspond to isolated polarons, while the formation energy of the exciton polaron (216\,meV) corresponds to a supercell with an exciton concentration at 7.0$\times$10$^{19}$~cm$^{-3}$ and is referenced to the lowest free exciton with zero momentum. 
}
\end{figure*}

\section*{Large polarons} The close agreement between our calculations and prior theoretical and experimental work on small polarons in \ch{TiO2} motivates us to proceed with the remaining members of the polaron family, namely hole polarons in rutile, electron polarons in anatase, and exciton polarons in both polymorphs. 
Inspection of Fig.~\ref{fig:small_polarons}(c) and Fig.~\ref{fig:small_polarons}(d) shows that both the valence band edge in rutile and the conduction band edge in anatase have a single anisotropic valley; therefore, before performing full-blown \textit{ab initio} calculations, we first estimate the size of the corresponding polarons using the semiclassical anisotropic Landau-Pekar model~\cite{guster2021frohlich}, which is derived in SI Appendix, Supplementary Note~3. We find that possible hole polarons in rutile and electron polarons in anatase would span at least 1.3\,nm and 4.3\,nm, respectively, ruling out the possibility of small polarons. These estimates are in line with the relatively light effective masses of holes in rutile (1.94\,$m_e$) and electrons in anatase (0.42\,$m_e$ in the $ab$ plane), and are consistent with prior experimental and theoretical studies which did not find such small polarons~\cite{yang2013intrinsic,berger2005light,zhang2007niobium,panayotov2012infrared,livraghi2011nature,setvin2014direct}. Exceptions to this consensus view are discussed in SI Appendix, Supplemental Note~4. An overview of the effective masses in these compounds is given in Fig.~\ref{fig:2d_e_polarons}(c).

While the anisotropic Landau-Pekar model points to the possibility of large polarons in \ch{TiO2}, these species have not been identified until now because they would require supercells containing many thousands of atoms, which is computationally prohibitive. Our methodology circumvents this limitation by solving the polaron problem in reciprocal space~\cite{sio2019ab, sio2019polarons,sio2023polarons,dai2023explrn_prb}, and allows us to take the first glimpse at large polarons in \ch{TiO2}.

In agreement with the anisotropic Landau-Pekar model, Fig.~\ref{fig:large_h_polarons}(a) shows that large hole polarons can exist in rutile for a hole density below the Mott metal-insulator transition at 1.7$\times$10$^{19}$\,cm$^{-3}$, which corresponds to one polaron per 1,000 rutile unit cells. In this regime, the polaron consists of a superposition of O-$2p$ orbitals modulated by a nearly isotropic Gaussian envelope of width 1.3\,nm, see Fig.~\ref{fig:large_h_polarons}(b). 
This quasiparticle predominantly arises from the interaction of holes with long-wavelength longitudinal optical phonons [SI Appendix, Fig.~S3(b)], therefore rutile hosts large Fr\"ohlich-type hole polarons. The formation energy of this polaron is 54~meV, but the migration barrier between adjacent unit cells is of only 0.07\,meV [Fig.~\ref{fig:large_h_polarons}(c)]. 
Therefore, this polaron will effectively behave as a free carrier in transport measurements, with a resistivity following the Bloch-Gr\"uneisen law, but with an effective mass enhanced by a factor 1.6 with respect to the band mass (see SI Appendix, Supplemental Note~5 for estimate using Feynman's model).

Fig.~\ref{fig:2d_e_polarons}(a) shows that anatase \ch{TiO2} hosts large quasi-two-dimensional electron polarons. These quasiparticles are stable below the Mott density of 4.4$\times$10$^{18}$\,cm$^{-3}$ (SI Appendix, Fig.~S4), and consist of Ti-$3d_{xy}$ orbitals modulated by an envelope function that spans 5\,nm along the $c$ axis but is delocalized in the $ab$ plane [Fig.~\ref{fig:2d_e_polarons}(b)]. The quasi-2D nature of this polaron is consistent with the anisotropic Landau-Pekar model shown in Fig.~\ref{fig:2d_e_polarons}(d) (see SI Appendix, Supplemental Note~3 for a detailed comparison between model and \textit{ab initio} calculations), and is a consequence of the highly anisotropic character of the conduction band bottom. This anisotropy reflects the fact that the underlying $d_{xy}$ orbitals have almost vanishing overlap along the $c$ axis. We note that, if we could perform calculations for even larger supercells, we would expect to find a very large but localized polaron, as in the model shown in Fig.~\ref{fig:2d_e_polarons}(d).

Similar low-dimensional polarons have been identified as metastable species in the related $d^0$ transition metal oxides \ch{HfO2} and \ch{ZrO2}~\cite{mckenna2012two}. Unlike in these other compounds, the large electron polaron in anatase \ch{TiO2} is stable at low temperature, with a formation energy of 10\,meV. 
This large electron polaron was originally identified through the observation of phonon replica bands in angle-resolved photoelectron spectroscopy (ARPES) maps on photo-doped single-crystal anatase \ch{TiO2}~\cite{PhysRevLett.110.196403,verdi2017origin}. To our knowledge, this is the first real-space visualization of the ARPES polaron of anatase \ch{TiO2}; this finding is of historical significance as the measurements of Ref.~\citenum{PhysRevLett.110.196403} marked the beginning of a series of ARPES investigations of polarons in many other complex oxides and 2D materials~\cite{wang2016tailoring,riley2018crossover,cancellieri2016polaronic,chen2018emergence,kang2018holstein,yukawa2016phonon}.

The quasi-2D nature of the large electron polaron in anatase within the $ab$ plane suggests that polaron formation should have no impact on in-plane carrier transport. To verify this hypothesis, we compute the electron mobility via the \textit{ab initio} Boltzmann transport equation without accounting for polaron formation, but including electron-phonon scattering and ionized-impurity scattering of free electrons~\cite{ponce2018towards, ponce2020first,leveillee2023ab, lu2019efficient, lu2022first,mustafa2016ab,li2015electrical}, {\color{black}as shown in Fig.~4(e) and SI Appendix, Fig.~S6.} Fig.~\ref{fig:2d_e_polarons}(e) shows that our calculations correctly reproduce the temperature dependence of the measured mobility, thereby confirming that electron polarons in anatase effectively behave as free carriers.

\section*{Exciton polarons} By solving Eqs.~(\ref{eqn:plrneqn})-(\ref{eqn:bmat_plrn}) we do not find any localized exciton polarons in rutile. This finding is consistent with the observation that the formation energies of electron polaron and hole polaron in rutile are very similar, 111\,meV and 54\,meV, respectively: Under these conditions, the strength of the electron-lattice and hole-lattice interactions tend to cancel each other out [see Eq.~(S1)], and the resulting exciton-phonon coupling is too weak to drive self-trapping~\cite{dai2023explrn_prb}. 
We note that potential signatures of localized excitons have been reported in the photoluminescence spectra of rutile~\cite{gallart2018temperature}, but they can be associated with excitons trapped at oxygen vacancy sites or surfaces as opposed to intrinsic exciton polarons.

The situation is very different in anatase, where the formation energies for the electron and the hole polarons differ by an order of magnitude (10\,meV vs.\ 312\,meV). In this case, the exciton-phonon coupling is sufficiently strong to drive the formation of exciton polarons, and we obtain stable polaronic species with a formation energy of 216\,meV as measured from the lowest free exciton state with zero momentum [Fig.~\ref{fig:exciton_polarons}(c)]. Fig.~\ref{fig:exciton_polarons}(a) shows the hole density and the electron density of this exciton polaron. The hole density is strongly localized along the $c$ axis, but more diffuse in the $ab$ plane as compared to the hole polaron in Fig.~\ref{fig:small_polarons}(b). Meanwhile, the electron density maintains the quasi-2D character of the electron polaron in Fig.~\ref{fig:2d_e_polarons}(a). Given that this exciton polaron is essentially a quasi-2D excitation nearly delocalized in the $ab$ plane, our calculations also rule out the presence of intrinsic STEs in anatase \ch{TiO2}. 
{\color{black}To further confirm this point, we have performed calculations of the exciton polaron hopping barriers, and we found the barier to be much smaller than thermal energies at room temperature, and much smaller than the hopping barriers for the small polarons, See SI Appendix, Fig. S5.}
Since the vertical ionization energy of the hole polaron in anatase  [1.37\,eV, Fig.~\ref{fig:small_polarons}(e)] is comparable to the experimental Stokes shift of 1.1\,eV~\cite{gallart2018temperature,watanabe2005time,tang1995urbach,hosaka1997excitonic}, proposed experimental observations of the STE in anatase might in fact correspond to the formation of independent hole polarons and electron polarons.

\begin{figure}
    \centering
    \includegraphics[width=1.0\columnwidth]{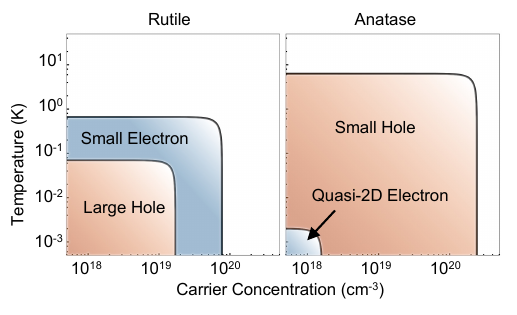}
    \caption{
     \textbf{Temperature-density phase diagrams of polarons in TiO$_2$.} Summary view of the existence regions of polarons in rutile and anatase TiO$_2$.
     The critical carrier density $n_{\rm c}$ is estimated at zero temperature from the extrapolated formation energy plots in SI Appendix, Fig. S4; the critical temperature $T_{\rm c}$ in the dilute limit is obtained by inverting $3/2 N k_{\rm B} T_{\rm c} = E_{\rm f}$, where $N$ is the critical Born-von-K\'arm\'an (BvK) supercell size, $E_{\rm f}$ is the extrapolated polaron formation energy, and $k_{\rm B}$ is the Boltzmann constant. Since we are currently unable to capture a fully localized electron polaron in anatase, in this case we determine the critical electron concentration using the width estimated from the anisotropic Landau-Pekar model, which corresponds to a 20$\times$20$\times$12 equivalent BvK supercell.
     The contours are guides to the eye. 
    \label{fig:phase_diagram}
    }
\end{figure}

\section*{Conclusion}
Fig.~\ref{fig:phase_diagram} presents a summary view of the temperature-density phase diagram of the polaronic species identified in this work. We see that both rutile and anatase \ch{TiO2} harbor a very rich variety of polaronic quasiparticles at low doping ($<$10$^{18}$\,cm$^{-3}$) and low temperature ($<$10\,K). In particular, the most exotic species, such as the quasi-2D polarons, should be the most difficult to probe given their weak formation energy.
The map outlined in Fig.~\ref{fig:phase_diagram} does not take into account more complex effects such as defect physics and polaron-polaron correlations, but it should serve as a guideline to design new experiments aimed at detecting large polarons and exciton polarons in \ch{TiO2}. Charting polaron maps of this type constitutes the first step toward quantitative  investigations of polaron physics and its potential applications in complex oxides and other quantum materials.

\section*{Methods}
\label{app:Methods}
{\color{black}The \textit{ab initio} theories of charged polarons and exciton polarons are derived from the variational minimization of the total energy of the polaronic system. For charged polarons, the formation energy is expressed as a functional of the single-particle polaron wave function and atomic displacements, as described in Ref.~\cite{sio2019ab}; for exciton polarons, the formation energy is expressed a functional of the two-particle exciton polaron wave function and atomic displacements, as discussed in Ref.~\cite{dai2023explrn_prb}. Minimization of these functionals with respect to wavefunctions and displacements yields the polaron solutions. In order to perform these calculations without resorting to computationally demanding supercells, we make the approximation of harmonic lattice dynamics and linear electron-phonon or exciton-phonon couplings; under these approximation, all variational parameters can be recast in a basis of Bloch waves (for electrons, excitons, or phonons) and the couplings can be computed from unit-cell calculations.
Specifically,} polarons and exciton polarons are expressed as coherent linear superpositions of Bloch states~\cite{sio2019ab, dai2023explrn_prb}:
\begin{align}
    &\psi^\mathrm{p}(\bmr) = \frac{1}{\sqrt{N}} \sum_{n\bmk} A^\mathrm{p}_{n\bmk} \psi_{n\bmk}(\bmr),
    \label{eqn:plrn_decomp}
    \\
    &\psi^\mathrm{ep}(\bmr_{\rm e}, \bmr_{\rm h}) = \frac{1}{\sqrt{N}} \sum_{s\bmqq} A^\mathrm{ep}_{s\bmqq} \phi_{s\bmqq}(\bmr_{\rm e}, \bmr_{\rm h}),
    \label{eqn:explrn_decomp}
\end{align}
where $\psi^\mathrm{p}(\bmr)$ is the wavefunction of a charged polaron; $\psi^\mathrm{ep}(\bmr_{\rm e}, \bmr_{\rm h})$ is the wavefunction of the exciton polaron in the electron ($\bmr_{\rm e}$) and hole ($\bmr_{\rm h}$) variables; $\psi_{n\bmk}$ is the Kohn-Sham (KS) eigenstate of the undistorted crystal, with electronic band index $n$ and crystal momentum $\bmk$; $\phi_{s\bmqq}$ is the Bethe-Salpeter equation (BSE) eigenstate of the undistorted crystal, with exciton band index $s$ and center-of-mass exciton momentum~$\bmqq$~\cite{rohlfing1998electron, deslippe2012berkeleygw}. $N$ is the number of crystal unit cells in the Born-von-K\'arm\'an supercell~\cite{sio2022unified}.

The coefficients $A^\mathrm{p}_{n\bmk}$ and $A^\mathrm{ep}_{s\bmqq}$ are determined variationally, by minimizing the energy functional of the polaron or the exciton polaron~\cite{sio2019ab, dai2023explrn_prb}.
This minimization amounts to the solution of the following coupled nonlinear eigenvalue problem:
\begin{align}
    \label{eqn:plrneqn}
    &\sum_{m'\bmkk'}
    \bigg[
    \varepsilon_{m\bmkk} \delta_{mm'} \delta_{\bmkk\bmkk'}
    -
    \frac{2}{N}\sum_{\nu} B_{\bmkk-\bmkk'\nu} G_{mm'\nu}(\bmkk', \bmkk-\bmkk')
    \bigg] 
    \nonumber \\ & \hspace{20pt}\times A_{m'\bmkk'}
    = \varepsilon A_{m\bmkk},  \\
    &B_{\bmq \nu}  
    = 
    \frac{1}{N \hbar \omega_{\bmq \nu}} 
    \sum_{\substack{mm' \bmkk'}}
    A^*_{m'\bmkk'} A_{m\bmkk'+\bmq}
    G^*_{mm'\nu}( \bmkk', \bmq), \nonumber \\[-10pt] \label{eqn:bmat_plrn}
\end{align}
where $(m, \bmkk, \varepsilon_{m\bmkk})$ are replaced by $(n, \bmk, \varepsilon^{\mathrm{KS}}_{n\bmk})$ for charged polarons, and by $(s, \bmqq, \varepsilon^{\mathrm{BSE}}_{s\bmqq})$ for exciton polarons. Similarly, the general interaction matrix elements $G_{mm'\nu}(\bmkk, \bmq)$ are replaced by the standard electron-phonon coupling matrix elements $g_{nn'\nu}(\bmk, \bmq)$ for charged polarons, and by the exciton-phonon coupling matrix elements $\mathcal{G}_{ss'\nu}(\bmqq, \bmq)$ for exciton polarons~\cite{giustino2017electron, chen2020exciton, antonius2022theory}.
The exciton-phonon coupling matrix element is given by Eq.~(S1) in SI Appendix, Supplemental Note~1.
In the above expressions, $\varepsilon^{\mathrm{KS}}_{n\bmk}$ and $\varepsilon^{\mathrm{BSE}}_{s\bmqq}$ are the eigenvalues of the KS equation and the BSE equation, respectively, and $\omega_{\bmq\nu}$ is the frequency of a phonon with wavevector $\bmq$ and branch $\nu$. These quantities are for the undistorted structure.
We use \textsc{Quantum ESPRESSO}~\cite{giannozzi2009quantum, giannozzi2017advanced,mostofi2014updated}, \textsc{EPW}~\cite{lee2023electron}, and \textsc{BerkeleyGW}~\cite{deslippe2012berkeleygw} to compute quantities needed in Eqs.~(\ref{eqn:plrneqn})-(\ref{eqn:bmat_plrn}).
Details on the computational setup, including the methodology for computing polaron displacement patterns, formation energies, potential energy landscape, charge density isosurfaces, and transport, are provided in SI Appendix, Supplemental Note~1.

\begin{acknowledgments}
This research was primarily supported by the Computational Materials Sciences Program funded by the US Department of Energy, Office of Science, Basic Energy Sciences, under award no. DE-SC0020129 (EPW development, calculations and analysis). Part of this research was supported by the National Science Foundation, Office of Advanced Cyberinfrastructure under Grant No. {2103991} of the Cyberinfrastructure for Sustained Scientific Innovation program, and the NSF Characteristic Science Applications for the Leadership Class Computing Facility program under Grant No. {2139536} (development of exciton polaron module).  This research used resources of the National Energy Research Scientific Computing Center and the Argonne Leadership Computing Facility, which are DOE Office of Science User Facilities supported by the Office of Science of the U.S. Department of Energy, under Contracts No. DE-AC02-05CH11231 and DE-AC02-06CH11357, respectively. The authors also acknowledge the Texas Advanced Computing Center (TACC) at The University of Texas at Austin for providing access to Frontera and Lonestar6 (http://www.tacc.utexas.edu).
\end{acknowledgments}

\bibliography{literature_v2}



\section*{Supplementary material}

# Identification of large polarons and exciton polarons in rutile and anatase polymorphs of titanium dioxide

## Supplemental Material

Zhenbang Dai[1,2] and Feliciano Giustino[1,2,*]

[1]*Oden Institute for Computational Engineering and Sciences,*
*The University of Texas at Austin, Austin, Texas 78712, USA*
[2]*Department of Physics, The University of Texas at Austin, Austin, Texas 78712, USA*

(Dated: October 22, 2024)

**Contents**

---

[*] fgiustino@oden.utexas.edu

**Supplemental Note 1: Extended description of computational methods**

To investigate polarons in $TiO_2$, we employ conventional unit cells for both rutile and anatase for ease of visualization; these cells contain 6 and 12 atoms, respectively. In the case of exciton polarons and Boltzmann transport calculations for anatase, we use the 6-atoms primitive unit cell to reduce computational cost. The initial structures are taken from from the Materials Project database [1]; both atomic coordinates and lattice vectors are subsequently optimized.

To compute optimized structure, Kohn-Sham (KS) states and energies, and phonon eigenmodes and frequencies, we perform density functional theory (DFT) and density functional perturbation theory (DFPT) calculations using the QUANTUM ESPRESSO package [2, 3]. We employ the local density approximation (LDA) to the DFT exchange and correlation functional, norm-conserving pseudopotentials [4, 5], and a planewaves kinetic energy cutoff of 90 Ry. The choice of the LDA is motivated by the fact that the PBE generalized gradient approximation (GGA) [6] yields soft phonons in rutile, while LDA phonon frequencies are in good agreement with experiments [7]. In general, one needs to first choose an appropriate exchange-correlation functional and converge the electronic and lattice dynamics calculations before moving to any polaron calculations. More discussions about the influence of the different exchange-correlation functionals on polaron calculations can be found in Supplemental Note 6.

We use EPW [8, 9] and WANNIER90 [10] to compute the electron-phonon coupling matrix elements, polarons [11], and to perform mobility calculations using the Boltzmann transport equation (BTE) [12–14]. BERKELEYGW is used to conduct GW/BSE (Bether-Salpeter equation) calculations with finite exciton momentum [15–17]. We set the kinetic energy cutoff for the dielectric matrix to 10 Ry; we include 24 valence bands and 276 conduction bands for rutile; and 24 valence bands and 476 conduction bands for anatase. The BSE kernel for rutile is constructed using 10 valence bands and 6 conduction bands; for anatase we employ 12 valence bands and 10 conduction bands.

We use a 6×6×6 uniform Brillouin-zone grid for DFT ground state calculations and to generate coarse-grid quantities needed for Wannier-Fourier interpolation (polarons and BTE). For BTE calculations of mobilities we use fine grids with 60×60×60 points, which are sufficient to achieve converged results. In the case of exciton polarons, we work directly with the coarse grid (i.e., without Wannier-Fourier interpolation) to ensure consistency between the phases of KS states between EPW and BERKELEYGW. For the rutile phase, we use grids of up to 8×8×8 points to sample the crystal momentum $\mathbf{k}$, the phonon wavevector $\mathbf{q}$, and the exciton momentum $\mathbf{Q}$, corresponding to a Born-von Kármán (BvK) supercell containing 3,072 atoms; however we could not find any localized exciton polaron in this case. For the anatase phase, we find a localized solution to the exciton-polaron equations using a 6×6×6 grid, corresponding to a supercell with 1,296 atoms.

Visualization of crystal structures, charge densities, and displacement patterns are performed using VESTA [18].

The exciton-phonon coupling matrix element appearing in Eqs. (3) and (4) of the main text is given by [19–22]:

$$\mathcal{G}_{ss'\nu}(\mathbf{Q},\mathbf{q}) = \sum_{vc\mathbf{k}} a_{vc\mathbf{k}}^{s\mathbf{Q}+\mathbf{q}*} \left[ \sum_{c'} g_{cc'\nu}(\mathbf{k}+\mathbf{Q},\mathbf{q}) a_{vc'\mathbf{k}}^{s'\mathbf{Q}} - \sum_{v'} g_{v'v\nu}(\mathbf{k},\mathbf{q}) a_{v'c\mathbf{k}+\mathbf{q}}^{s'\mathbf{Q}} \right]. \tag{S1}$$

Here, $a_{vc\mathbf{k}}^{s\mathbf{Q}}$ is the eigenvector of the BSE, and $g_{nn'\nu}(\mathbf{k},\mathbf{q})$ is the standard electron-phonon coupling matrix element. Both quantities are evaluated for the undistorted structure.

To compute the atomic displacements associated with polarons and exciton polarons, we proceed as follows. The equations below are summarized from Refs. 21, 23. Once the solution vectors $B_{\mathbf{q}\nu}$ are obtained, we determine the displacements using:

$$\Delta\tau_{\kappa\alpha p} = -\frac{2}{N} \sum_{\mathbf{q}\nu} B_{\mathbf{q}\nu} \left(\frac{\hbar}{2M_\kappa \omega_{\mathbf{q}\nu}}\right)^{1/2} e_{\kappa\alpha,\nu}(\mathbf{q}) e^{i\mathbf{q}\cdot\mathbf{R}_p}, \tag{S2}$$

where $\mathbf{q}$ is the phonon wavevector, $N$ is the number of unit cells in the BvK supercell, $M_\kappa$ is the mass of atom $\kappa$, $e_{\kappa\alpha,\nu}(\mathbf{q})$ is polarization vector of the phonon with momentum $\mathbf{q}$, branch $\nu$, and frequency $\omega_{\mathbf{q}\nu}$; $\mathbf{R}_p$ is the lattice vector of the $p$-th unit cell in the BvK supercell [24].

The formation energy of the polaron and the exciton polarons obtained from the solutions of Eqs. (3) and (4) of the main text are given by [21, 23]:

$$\Delta E_\mathrm{p} = \frac{1}{N} \sum_{n\mathbf{k}} \left|A_{n\mathbf{k}}^\mathrm{KS}\right|^2 (\varepsilon_{n\mathbf{k}}^\mathrm{KS} - \varepsilon_\mathrm{CBM}^\mathrm{KS}) - \frac{1}{N} \sum_{\mathbf{q}\nu} |B_{\mathbf{q}\nu}|^2 \hbar\omega_{\mathbf{q}\nu},$$

$$\Delta E_\mathrm{xp} = \frac{1}{N} \sum_{s\mathbf{Q}} \left|A_{s\mathbf{Q}}^\mathrm{BSE}\right|^2 (\varepsilon_{s\mathbf{Q}}^\mathrm{BSE} - \varepsilon_0^\mathrm{BSE}) - \frac{1}{N} \sum_{\mathbf{q}\nu} |B_{\mathbf{q}\nu}|^2 \hbar\omega_{\mathbf{q}\nu}, \tag{S3}$$

where $\varepsilon_{\text{CBM}}^{\text{KS}}$ is the energy of the conduction band minimum (CBM), and $\varepsilon_0^{\text{BSE}}$ is the energy of the lowest-lying exciton with zero momentum.

For the calculation of polaron hopping barriers, we compute the ground-state energy of the polaron at fixed atomic coordinates. In this case, only Eq. (3) of the main text is solved at fixed $B_{\mathbf{q}\nu}$. The energy is given by:

$$E_{\text{tot}} = \varepsilon + \frac{1}{N} \sum_{\mathbf{q}\nu} |B_{\mathbf{q}\nu}|^2 \hbar \omega_{\mathbf{q}\nu}, \tag{S4}$$

where $\varepsilon$ is the eigenvalue of the polaron or exciton-polaron equation. The vector $B_{\mathbf{q}\nu}$ is determined by setting the atomic displacements $\Delta\tau_{\kappa\alpha p}$ and inverting Eq. (S2).

In order to analyze polarons in terms of their underlying KS states and phonons, we use the spectral functions [23]:

$$A^2(E) = \frac{1}{N} \sum_{n\mathbf{k}} |A_{n\mathbf{k}}^{\text{KS}}|^2 \delta(E - \varepsilon_{n\mathbf{k}}^{\text{KS}} + \varepsilon_{\text{CBM}}^{\text{KS}}), \tag{S5}$$

$$B^2(E) = \frac{1}{N} \sum_{\mathbf{q}\nu} |B_{\mathbf{q}\nu}|^2 \delta(E - \hbar\omega_{\mathbf{q}\nu}). \tag{S6}$$

To plot the partial electron or hole charge densities associated with exciton polarons, we use the following expressions [21]:

$$n_e(\mathbf{r}_e) = \int d\mathbf{r}_h |\Psi(\mathbf{r}_e, \mathbf{r}_h)|^2 = \frac{1}{N} \sum_{v\mathbf{k}} |L_{v\mathbf{k}}(\mathbf{r}_e)|^2, \tag{S7}$$

$$n_h(\mathbf{r}_h) = \int d\mathbf{r}_e |\Psi(\mathbf{r}_e, \mathbf{r}_h)|^2 = \frac{1}{N} \sum_{c\mathbf{k}} |L_{c\mathbf{k}}(\mathbf{r}_h)|^2, \tag{S8}$$

where the auxiliary functions $L_{v\mathbf{k}}$ and $L_{c\mathbf{k}}$ are given by:

$$L_{v\mathbf{k}}(\mathbf{r}_e) = \sum_{s\mathbf{Q}} \sum_c A_{s\mathbf{Q}} a_{vc\mathbf{k}}^{s\mathbf{Q}} \psi_{c\mathbf{k}+\mathbf{Q}}(\mathbf{r}_e),$$

$$L_{c\mathbf{k}}(\mathbf{r}_h) = \sum_{s\mathbf{Q}} \sum_v A_{s\mathbf{Q}} a_{vc\mathbf{k}-\mathbf{Q}}^{s\mathbf{Q}} \psi_{v\mathbf{k}-\mathbf{Q}}^*(\mathbf{r}_h). \tag{S9}$$

$$\tag{S10}$$

**Supplemental Note 2: Polaron hopping transport**

From the potential energy landscape calculated using Eq. (S4), we obtain the hopping barrier $\Delta E_{\text{hop}}$, as shown in Fig. S2. These barriers are used to estimate the polaron mobility using the Emin-Holstein-Austin-Mott (EHAM) theory [25, 26]. The Emin-Holstein-Austin-Mott model assumes that the transport occurs via discrete hopping events between neighboring lattice sites. The rate for each event is determined by the frequency of hopping attempts, which is approximated as the characteristic vibration energy of the polaron, and the quantum-mechanical probability for intersite electron transfer. It can be shown that this probability is proportional to the adiabatic hopping barrier when there is no level crossing [27]. In the case of the electron polaron in rutile, there is no level crossing, therefore the model requires the hopping barrier $\Delta E_{\text{hop}}$, the frequency of the optical phonon mode that assists the hopping $\omega$, and the geometry of the host system that determines the number of sites $n$ accepting the polaron. These quantities are obtained from our polaron calculations. In this model, the rate of polaron transfer at a given temperature $T$ is given by:

$$k_{\text{et}} = \frac{\omega}{2\pi} \exp\left(\frac{-\Delta E_{\text{hop}}}{k_{\text{B}} T}\right), \tag{S11}$$

where $\omega$ is the characteristic frequency of optical phonons, and and $k_{\text{B}}$ is the Boltzmann constant. We take $\hbar\omega = 98.8$ meV based on Fig. S3(a) and (b), which corresponds to the frequency of the contributing LO mode near $\Gamma$ point. From the transfer rate $k_{\text{et}}$, the diffusion coefficient is calculated using:

$$D = R^2 n k_{\text{et}}, \tag{S12}$$

where $R$ is the distance between transfer sites and $n$ is the so-called the geometric factor and encodes the information on the hopping path. Following previous work on rutile TiO$_2$ [25–27], we take $n = 1$. The mobility is then obtained from the diffusion coefficient $D$ using Einstein's relation:

$$\mu = \frac{eD}{k_B T} \tag{S13}$$

where $e$ is the electron charge. In order to estimate the resistivity [13], we also need the carrier concentrations $n_c$, which we take from experiments [28]:

$$\rho = (en_c\mu)^{-1}. \tag{S14}$$

$$\tag{S15}$$

**Supplemental Note 3: Anisotropic Landau-Pekar model**

The Landau-Pekar model of polarons considers isotropic systems [23, 29–31]. This model was generalized to study systems with degenerate and anisotropic bands [32]. Here, we present this generalization with a focus on materials with significant anisotropy between the in-plane and the out-of-plane directions, in order to describe systems like anatase and rutile TiO$_2$. To this end, we generalize the kinetic energy to incorporate an in-plane effective mass $m_\parallel^*$ (in the $ab$ plane) and an out-of-plane effective mass $m_\perp^*$ (along the $c$ axis). We do not include dielectric anisotropy since both the static permitivity tensors $\epsilon^0$ and high-frequency permittivity tensors $\epsilon^\infty$ are diagonal and with components of $\kappa = (1/\epsilon^\infty - 1/\epsilon^0)^{-1}$ only 2% and 11% deviation from their isotropic average for anatase and rutile, respectively.

The generalized Landau-Pekar energy functional of the polaron wavefunction $\psi$ is:

$$E_{\rm LP}[\psi] = -\frac{\hbar^2}{2} \int d\mathbf{r}\psi^*(\mathbf{r})\left(\frac{1}{m_\parallel^*}\frac{\partial^2}{\partial x^2} + \frac{1}{m_\parallel^*}\frac{\partial^2}{\partial y^2} + \frac{1}{m_\perp^*}\frac{\partial^2}{\partial z^2}\right)\psi(\mathbf{r}) - \frac{1}{2}\frac{e^2}{4\pi\epsilon_0}\frac{1}{\kappa}\int d\mathbf{r}d\mathbf{r}'\frac{|\psi(\mathbf{r})|^2|\psi(\mathbf{r}')|^2}{|\mathbf{r}-\mathbf{r}'|}. \tag{S16}$$

We minimize this functional variationally using the normalized Gaussian trial wavefunction:

$$\psi(\mathbf{r}) = \frac{(2\pi)^{-3/4}}{\sqrt{\sigma_\parallel^2 \sigma_\perp}} \exp\left(-\frac{x^2+y^2}{4\sigma_\parallel^2}\right) \exp\left(-\frac{z^2}{4\sigma_\perp^2}\right), \tag{S17}$$

where $\sigma_\parallel$ and $\sigma_\perp$ represent the in-plane and out-of-plane spreads of the polaron, respectively, and serve as variational parameters. By inserting Eq. (S17) inside (S16) we obtain:

$$\frac{E_{\rm LP}}{\hbar\omega_{\rm LO}} = \left[\frac{1}{2(m_\parallel^*/m^*)(\sigma_\parallel/\sigma_\perp)^2} + \frac{1}{4(m_\perp/m^*)}\right]\frac{1}{(\sigma_\perp/l)^2} - \frac{1}{\sqrt{\pi}}\frac{\alpha}{(\sigma_\perp/l)}\frac{\tan^{-1}\sqrt{(\sigma_\parallel/\sigma_\perp)^2-1}}{\sqrt{(\sigma_\parallel/\sigma_\perp)^2-1}}, \tag{S18}$$

having defined the average mass:

$$\frac{1}{m^*} = \frac{1}{3}\left(\frac{2}{m_\parallel^*} + \frac{1}{m_\perp^*}\right), \tag{S19}$$

the characteristic polaron size [33]:

$$l = \sqrt{\frac{\hbar}{m^*\omega_{\rm LO}}}, \tag{S20}$$

and the average coupling strength [33]:

$$\alpha = \frac{e^2}{4\pi\epsilon_0\hbar}\sqrt{\frac{m^*}{2\hbar\omega_{\rm LO}}}\frac{1}{\kappa}. \tag{S21}$$

In all these expressions, $\omega_{\rm LO}$ is the characteristic frequency of the longitudinal optical phonon. If we set $m_\parallel^* = m_\perp^*$ as well as $\sigma_\parallel = \sigma_\perp$ in Eq. (S18), and minimize with respect to the spread, we recover the standard solution [33] $E_{\rm LP,min} = -\alpha^2\hbar\omega_{\rm LO}/3\pi$.

In the anisotropic case of TiO$_2$, we carry out the minimization of Eq. (S18) with respect to both $\sigma_\parallel$ and $\sigma_\perp$. In particular, for the electron polaron in anatase, we use $\kappa = 7.96$, $m_\parallel^* = 0.42\,m_0$ (where $m_0$ is the free electron mass), $m_\perp^* = 3.96\,m_0$, and $\hbar\omega_{\rm LO} = 90.3$ meV. The polaron solution corresponds to:

$$E_{\rm LP,min} = -17 \text{ meV}, \qquad \sigma_\parallel = 18.1 \text{ Å (FWHM = 4.3 nm)}, \qquad \sigma_\perp = 8.2 \text{ Å (FWHM = 1.9 nm)}, \tag{S22}$$

where FWHM stands for full width at half maximum, and is evaluated as $2\sqrt{2\log 2}\,\sigma$. For the hole polaron in rutile, we use $\kappa = 8.18$, $m_\parallel^* = 1.94\,m_0$, $m_\perp^* = 2.85\,m_0$, and $\hbar\omega_{\rm LO} = 98.8$ meV. The polaron solution corresponds to:

$$E_{\rm LP,min} = -47 \text{ meV}, \qquad \sigma_\parallel = 5.4 \text{ Å (FWHM = 1.3 nm)}, \qquad \sigma_\perp = 4.8 \text{ Å (FWHM = 1.1 nm)}. \tag{S23}$$

Given the simplicity of the Landau-Pekar model, these values are in surprising agreement with our full-blown *ab initio* calculations. For example, in the case of anatase we find a formation energy of 10 meV and a FWHM along the *c* axis of 5 nm, which are of the same magnitude as the values 17 meV and 1.9 nm in Eq. (S22). Similarly, in the case of rutile our *ab initio* calculations yield a formation energy of 54 meV and a nearly-isotropic spread of 1.3 m, which match the data in Eq. (S23) of 47 meV and 1.1–1.3 nm, respectively.

### Supplemental Note 4: Previous work on small polarons

Both our *ab initio* calculations and the anisotropic Landau-Pekar model support the notion that rutile TiO$_2$ hosts large hole polarons, and anatase TiO$_2$ hosts large electron polarons. These findings are consistent with several previous studies [28, 34–38]. However, there are a few reports that propose the presence of small polarons rather than large polarons in these cases [25, 27, 39–41]. Here, we reconcile these conflicting findings.

Experimentally, there is one electron paramagnetic resonance (EPR) study where the authors report a small self-trapped hole polaron in the rutile phase at 4 K by analyzing the angular-dependence of an EPR signal [39]. However, the detection of such an EPR signal requires continuous illumination of the TiO$_2$ sample, and the signal vanishes once the temperature increases to 10 K. These observations are more aptly explained by a weakly bound mobile hole polaron such as the one identified in the present study. In fact, the disappearance of the EPR signal at 10 K (0.8 meV) is consistent with the small migration barrier in Fig. 3(c), and suggests that, at this temperature, the polaron migrates like a free-carrier, and is more difficult to detect via EPR. Furthermore, the need for continuous illumination suggests that these polarons are mobile species that diffuse away on fast timescales, making their detection via EPR more difficult once light is turned off. Therefore, experimental data of Ref. 39 are compatible with our finding of large hole polarons in rutile.

On the theory side, Ref. 25 and 27 find a small hole polaron in rutile and a small electron in anatase, respectively, using Hubbard-corrected DFT+$U$. However, in these studies the effective Hubbard parameter $U$ is chosen to be unrealistically large (10 eV). For example, typical $U$ values for TiO$_2$ are in the range 2.5-4.5 eV [42], and the Hubbard parameters determined *ab initio* from the constrained random-phase approximation are 3.9 eV and 4.1 eV for rutile and anatase, respectively [43]. Using very large values of $U$ leads to artificial electron localization as a result of the overcompensation for the DFT self-interaction error [38, 44]. Similar effects are also observed with hybrid functional calculations in anatase, where by artificially increasing the fraction of exact exchange one finds small electron polarons [40, 41, 45]. Furthermore, due to computational cost, only relatively small supercells could be tested in prior work, thereby precluding the possibility of observing large polarons. The approach used in the present work overcomes these limitations; in particular, the self-interaction is removed exactly [23], therefore the use of Hubbard-corrected DFT or the tuning of the exchange fraction in hybrid functionals are not needed; additionally, the reciprocal space formulation allows us to explore very large supercells that are needed to enclose the wavefunctions of large polarons.

### Supplemental Note 5: Polaronic mass enhancement from Feynman's model

We estimate the polaronic mass enhancement in TiO$_2$ using the Feynman all-coupling model [46, 47]. In this model, the formation energy and the effective mass of the polarons depend on the characteristic phonon frequency $\omega_{\rm LO}$ and the effective coupling strength $\alpha$ (as in the Landau-Pekar model discussed in Suppemental Note 3). The formation energy is obtained by minimizing the following function

$$\frac{\Delta E_{\rm f}}{\hbar\omega_{\rm LO}} = \frac{3}{2}(v-w) - \frac{3}{4}\frac{v^2-w^2}{v} - \alpha\pi^{-\frac{1}{2}}\frac{v}{w}\int_0^\infty d\tau\, e^{-\tau}[\tau j(\tau)]^{-\frac{1}{2}} \tag{S24}$$

with respect to the parameters $v$ and $w$. The auxiliary function $j(\tau)$ in the above expression is given by:

$$j(\tau) = 1 + \frac{v}{w^2}\left(1 - \frac{w^2}{v^2}\right)\frac{1}{\tau}(1 - e^{-v\tau}). \tag{S25}$$

The Feynman polaron effective mass $m_{\rm F}^*$ is obtained by using in the following expression the values $v$ and $w$ that minimize the formation energy:

$$m_{\rm F}^* = m^* \left[ 1 + \frac{1}{3}\alpha\pi^{-\frac{1}{2}} \frac{v^3}{w^3} \int_0^\infty d\tau e^{-\tau} [\tau j(\tau)]^{\frac{1}{2}} \right], \tag{S26}$$

where $m^*$ is the band effective mass.

Here, we extract the effective isotropic coupling strength $\alpha$ of rutile and anatase $TiO_2$ using the average mass in Eq. (S19), and insert this value inside Eq. (S21). The characteristic phonon frequencies are the same as in Supplemental Note 3. We find $\alpha = 2.12$ for rutile, and $\alpha = 1.19$ for anatase. Our calculated KS effective masses are:

$$\text{Rutile TiO}_2, \text{ holes}: \qquad m_\parallel^* = 1.94\,m_0, \qquad m_\perp^* = 2.85\,m_0, \tag{S27}$$

$$\text{Anatase TiO}_2, \text{ electrons}: \qquad m_\parallel^* = 0.42\,m_0, \qquad m_\perp^* = 3.96\,m_0. \tag{S28}$$

The corresponding effective masses enhanced by polaronic effects in the Feynman model are:

$$\text{Rutile TiO}_2, \text{ holes}: \qquad m_\parallel^* = 3.13\,m_0, \qquad m_\perp^* = 4.60\,m_0, \tag{S29}$$

$$\text{Anatase TiO}_2, \text{ electrons}: \qquad m_\parallel^* = 0.53\,m_0, \qquad m_\perp^* = 4.98\,m_0. \tag{S30}$$

By comparing Eqs. (S27)-(S30), we find that the effective masses in rutile are enhanced by a factor 1.6, while the effective masses in anatase are enhanced by 1.2. The moderate mass enhancement in anatase justifies the use of the BTE for calculating the carrier mobility in this polymorph.

**Supplemental Note 6: Influence from different exchange-correlation functionals**

The lattice dynamics of rutile $TiO_2$ is known to be sensitive to the exchange-correlation functionals, and it has been shown that the PBE functional gives rise to a spurious ferroelectric phonon mode at $\mathbf{q} = 0$ with imaginary frequency, causing an artificial symmetric breaking [7]. Thus, the PBE functional is inadequate to study polarons in the rutile phase of $TiO_2$. Similarly, the PBEsol functional yields phonon dispersions without soft modes, but it underestimates several zone-center optical modes as compared to experiments [6,47]. For these reasons, we have employed the LDA functional in our calculations; this functional yields accurate phonon dispersions in rutile as compared to experiments.

If we employ the PBEsol functional [48] to perform the polaron calculations, the inaccuracy in the phonon dispersions, which is also reflected in the screening entering the electron-phonon matrix elements, causes a overestimation of the formation energy of the electron polaron and hole polaron, as we show in Table S2. Moreover, the tendency to a ferroelectric instability in both PBE and PBEsol lead to a less symmetric wavefunction for the electron polaron, as we show in Fig. S7(a).

In the case of anatase $TiO_2$, we are unaware of experimental phonon dispersions that we could compare with, but the phonon dispersions computed from LDA, PBE, and PBEsol do not show imaginary frequency modes. PBE and PBEsol still underestimate the phonon frequencies as compared with the LDA, but the consistent shape of phonon dispersions from all three functionals suggest that the polaron calculations should be less sensitive to the exchange-correlation functionals. Indeed, polaron calculations using these two functionals yield the same qualitative picture, while the formation energies differ by up to 42%, as shown in Table S2 and Fig. S7(b).

To the best of our knowledge, density-functional perturbation theory calculations with hybrid functionals are not implemented yet in any existing DFT software package, therefore at present we are unable to assess hybrid functionals.

Generally, while we expect numerical differences in the polaron energetics and wave functions when using hybrids, the qualitative aspects of the solution (localization/delocalization) should remain as in the LDA. We also emphasize that the present methodology removes the DFT self-interaction error from the outset, therefore the well-known sensitivity of polaron energetics to the Hartree self-interaction of the DFT functional is eliminated in our approach.

In summary, the conclusions in this manuscript are robust with respect to the choice of functional, and inspecting phonon dispersions before proceeding to polaron calculations is a good strategy to avoid artifact and obtain reliable solutions.

7| Polymorph | Polaron type | Formation energy | In-plane FWHM | Out-of-plane FWHM |
|---|---|---|---|---|
| Rutile | electron | 111 meV | 0.2 nm | 0.4 nm |
| | hole | 54 meV | 1.3 nm | 1.2 nm |
| | exciton | – | – | – |
| Anatase | electron | 10 meV | – | 5.0 nm |
| | hole | 312 meV | 0.2 nm | 0.3 nm |
| | exciton | 216 meV | 4.3 nm (e) <br> 1.8 nm (h) | 1.2 nm (e) <br> 0.6 nm (h) |

TABLE S1. **Summary of polaron energetics and size**. Calculated polaron formation energy and FHWM for rutile and anatase TiO$_2$. The formation energy of charged polarons correspond to the limit of isolated polarons (infinitely large BvK supercell); the formation energy of the exciton polaron is for a 6×6×6 BvK primitive supercell, which corresponds to an exciton density of $7.0 \times 10^{19}$ cm$^{-3}$.

| Polaron species | Polaron formation energy (meV) | | |
|---|---|---|---|
| | LDA | PBE | PBEsol |
| Rutile electron | 111 | - | 223 |
| Rutile hole | 54 | - | 86 |
| Anatase electron | 10 | 8 | 7 |
| Anatase hole | 312 | 443 | 370 |

TABLE S2. **Comparison of the polaron formation energies results from LDA and GGA functionals.** For the rutile phase, the PBE functional will give a spurious ferroelectric phonon mode at $\mathbf{q} = 0$ with imaginary frequency, causing an artificial symmetry breaking. Thus, the PBE functional is inadequate to study polarons in the rutile phase of TiO$_2$.

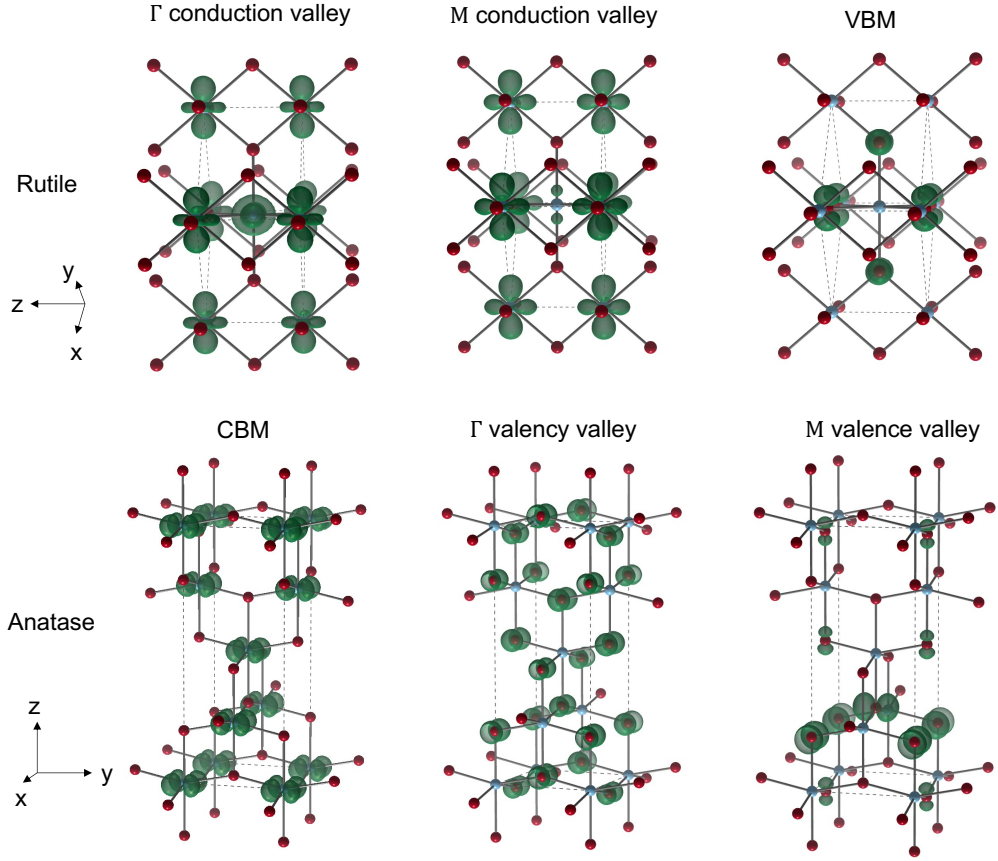

FIG. S1. **Orbital character of frontier states in rutile and anatase TiO$_2$**. Isosurfaces of the square moduli of the wavefunctions (dark green) for the valence band minimum (VBM) and conduction band maximum (CBM) states of rutile TiO$_2$ and anatase TiO$_2$. Since the conduction bands of rutile and the valence bands of anatase exhibit multiple valleys [Figs. 2(c) and 2(d) of the main text], we show the CBM and VBM states for each valley.

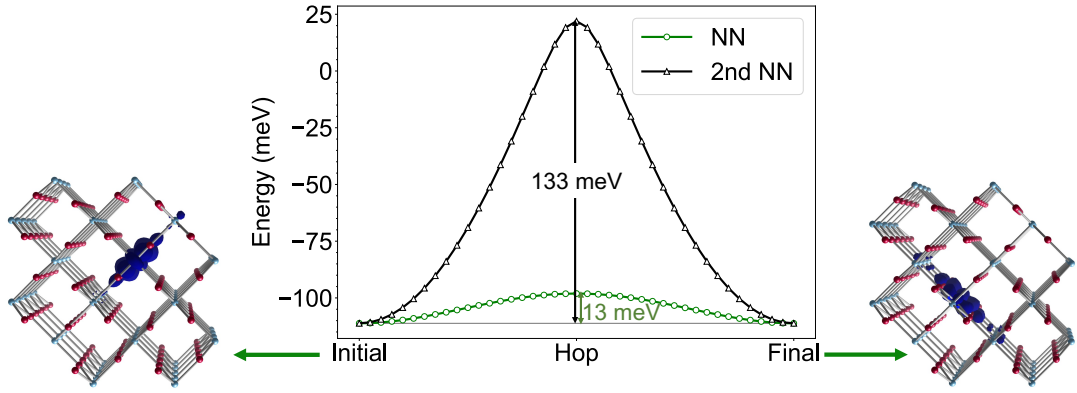

FIG. S2. **Polaron hopping barrier in rutile TiO$_2$**. Potential energy surface for hopping of the small electron polaron in rutile between nearest-neighbor (NN) and next-nearest-neighbor (2nd NN) sites. The hopping barriers are given by the height of the curves and are $\Delta E_{\rm hop} = 13\,{\rm meV}$ and $133\,{\rm meV}$ for NN and 2nd NN hops, respectively. The energy surface is obtained by calculating the polaron energy at fixed atomic coordinates. The coordinates are generated as a linear superposition between the polaronic distortion in the initial state and the final state, where the corresponding polaron charge density isosurfaces are displayed on the left and right, respectively.

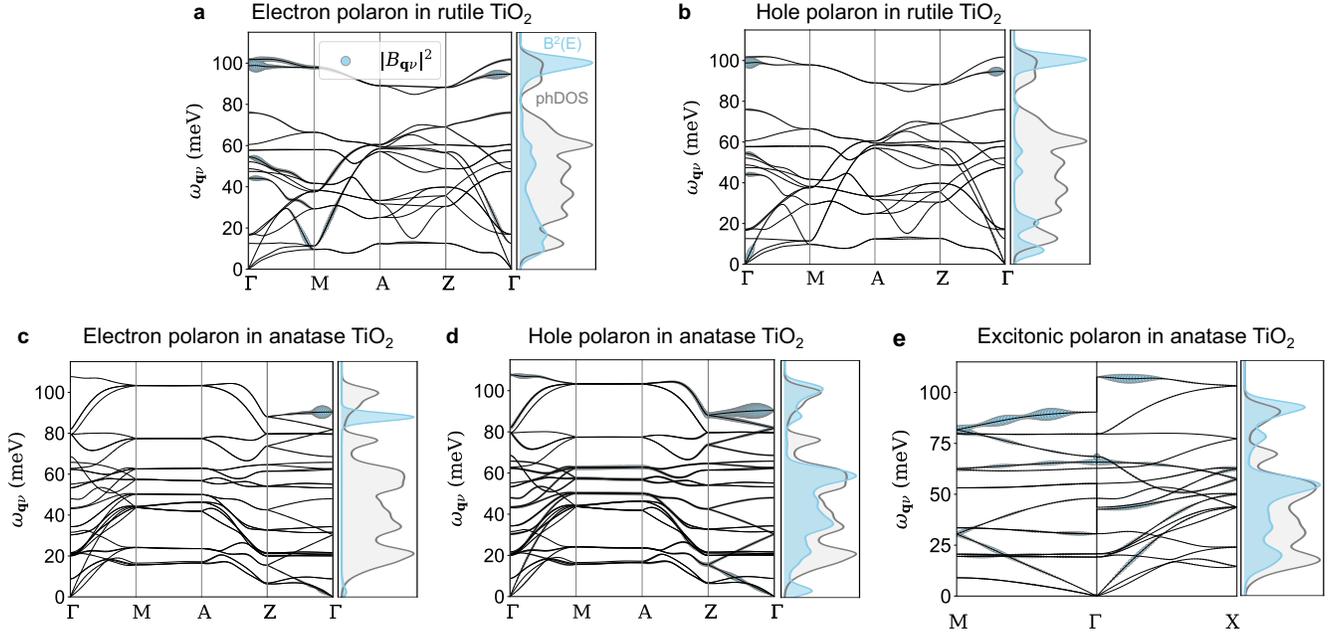

FIG. S3. **Phonon contributions to polarons in TiO$_2$**. Spectral decomposition of polarons into individual phonon modes. **a** Phonon dispersions in rutile, with $|B_{\mathbf{q}\nu}|^2$ for the electron polaron superimposed. **b** Same, for the hole polaron in rutile. **c,d**, **e** Same, for the electron, hole, and exciton polarons in anatase, respectively. The right panel in each subfigure displays the corresponding phonon spectral function $B^2(E)$ from Eq. (S6); these curves are normalized so that the top of the range coincides with the highest peak in each case. In the case of large charged polarons (hole polaron in rutile and electron polaron in anatase), the electron-phonon coupling is dominated by a single longitudinal optical phonon; this explains why the anisotropic Landau-Pekar model of Supplemental Note S3 works remarkably well in these cases.

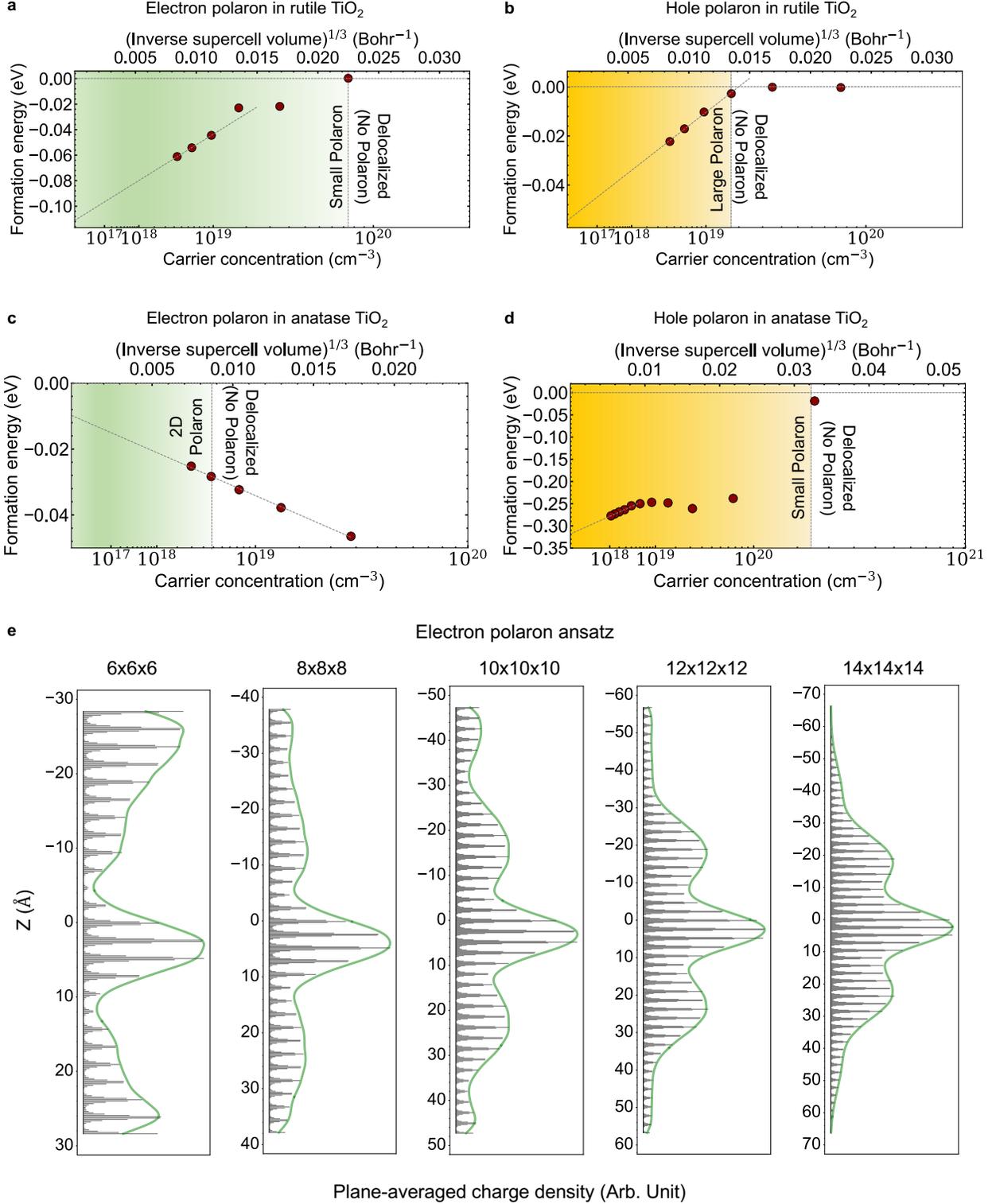

FIG. S4. **Polaron formation energies in $TiO_2$ vs. carrier concentration**. **a,b** Formation energy of electron polaron and hole polaron in rutile vs. density, respectively. **c,d** Formation energy of electron polaron and hole polaron in anatase, respectively. The extrapolation to vanishing density corresponds to the limit of a single polaron in the crystal, and is performed using linear regression on the three leftmost data points. The data points are plotted according to the formation energies and the inverse supercell lengths, and the carrier concentrations are computed separately for each supercell size. We note that, in the case of the electron polaron in anatase, a stable solution is found at all concentrations tested here. However, inspection of individual wavefunctions shows that some of these solutions do not correspond to polarons, but to charge density waves. To illustrate this point, in **e** we plot the planar-average of the electron polaron density along the $c$ axis of the anatase lattice for various BvK supercells. It is seen that the polaron forms only at densities below $4.4 \times 10^{18}\,\text{cm}^{-3}$, which corresponds to a $12 \times 12 \times 12$ BvK supercell.

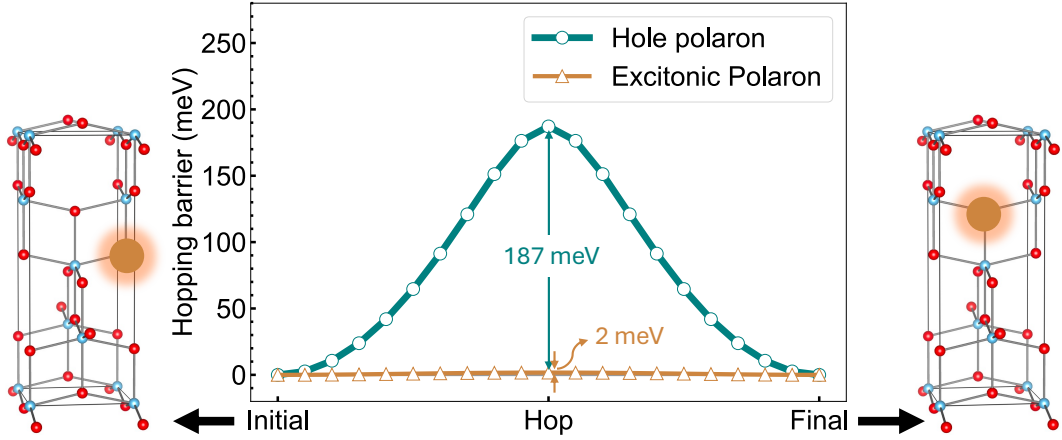

FIG. S5. **Hopping barriers of the exciton polaron and the hole polaron in anatase TiO$_2$ between the nearest neighbors.** The center of the exciton polarons or the hole polarons are marked as the orange circle, and their positions for the initial state and final state are displayed on the left and right, respectively. The hopping barrier for the exciton polaron for the nearest neighbors is much smaller than thermal energies at room temperature and also 90× smaller than the hopping barrier of the small hole polaron in anatase. We expect that this small hopping barrier does not lead to any significant self-trapping of exciton polarons.

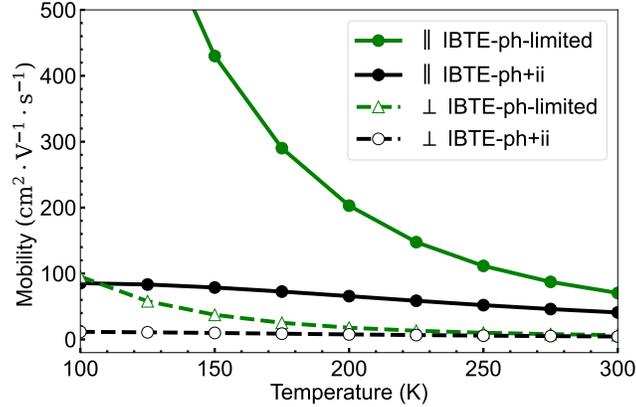

FIG. S6. **Electron mobility of anatase TiO$_2$ for both in-plane and out-of-plane directions.** Calculated via the *ab initio* Boltzmann transport equation, including both phonon scattering (BTE ph) and ionized-impurity scattering (BTE ph+ii). The room temperature in-plane and out-of-plane mobilities of anatase TiO$_2$, including both phonon and impurity scattering, are 40 cm$^2$/Vs and 4 cm$^2$/Vs, respectively.


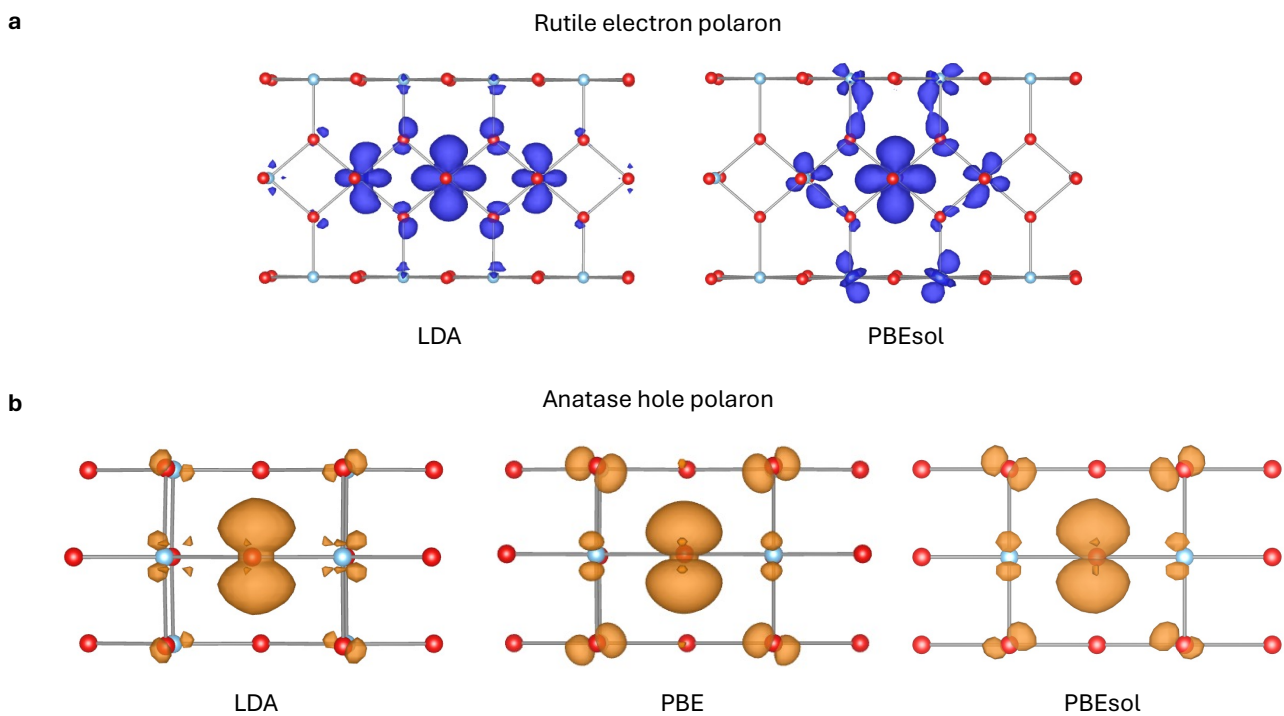

FIG. S7. **Comparison of the polaron wave functions from LDA and GGA functionals.** **a** The comparison of the rutile electron polaron wave function from LDA and PBEsol. **b** The comparison of the anatase hole polaron wave functions from LDA, PBE, and PBEsol.



\end{document}